\documentclass[11pt,psamsfonts]{amsart}%
\usepackage{amsmath}
\usepackage{amsthm}
\usepackage{amssymb}
\usepackage{amscd}
\usepackage{amsfonts}
\usepackage{amsbsy}
\usepackage{graphicx}
\usepackage[dvips]{psfrag}%
\newcommand{\bbox}{\ \hfill\rule[-1mm]{2mm}{3.2mm}}

\newtheorem {theorem} {Theorem}

\newtheorem {remark} {Remark}
\newtheorem {note} {Note}
\begin{document}
\title[Dynamics of a dumbbell satellite...]{Dynamics of a dumbbell satellite under the zonal harmonic effect of an oblate body}
\author[E.I. Abouelmagd, J.L.G. Guirao, J.A. Vera]{Elbaz I. Abouelmagd$^{1,2}$, Juan L.G. Guirao$^{3}$ and Juan A. Vera$^{4}$}

\address{$^{1}$ Mathematics Department, Faculty of Science and Arts (Khulais), King Abdulaziz,
University, Jeddah, Saudi Arabia}

\address{$^{2}$ Nonlinear Analysis and Applied Mathematics Research Group (NAAM)
Department of Mathematics, King Abdulaziz University, Jeddah, Saudi Arabia}

\email{eabouelmagd@gmail.com, eabouelmagd@kau.edu.sa}

\address{$^{3}$ Departamento de Matem\'{a}tica Aplicada y Estad\'{\i}stica. Universidad
Polit\'{e}cnica de Cartagena, Hospital de Marina, 30203--Cartagena, Regi\'{o}n
de Murcia, Spain.--Corresponding Author--}

\email{juan.garcia@upct.es}

\address{$^{4}$ Centro Universitario de la Defensa. Academia General del Aire.
Universidad Polit\'{e}cnica de Cartagena, 30720-Santiago de la Ribera,
Regi\'{o}n de Murcia, Spain.}

\email{juanantonio.vera@cud.upct.es}



\maketitle

\begin{abstract}
The aim of the present paper is to study the dynamics of a dumbbell satellite moving in a gravity field generated by an oblate body considering the effect of the zonal harmonic parameter. We prove that the pass trajectory of the mass center of the system is periodic and different from the classical one when the effect of the zonal harmonic parameter is non zero.
Moreover, we complete the classical theory showing that the equations of motion in the satellite approximation can be reduced to Beletsky's equation when the zonal harmonic parameter is zero. The main tool for proving these results is the Lindstedt--Poincare's technique.
\end{abstract}

\section{Introduction}
From the end of the sixth decade of the last century, a part of the mathematical community, has directed its attention to the study the so called \emph{dumbbell body or satellite} in central gravity, see for instance Mor\'{a}n \cite{Moran}, Schechter \cite{Schechter}, Brereton and Modi \cite{Brereton}, Beletsky \cite{Beletsky1, Beletsky2}, Maciejewski et al. \cite{Maciejewski};
 Kirchgraber et al. \cite{Kirchgraber},  Krupa et al. \cite{Krupa}, Elipe et
al. \cite{Elipe},  Burov and Dugain \cite{Burov1} or Nakanishi et al. \cite{Nakanishi}.

\medskip

Recall that a \emph{dumbbell body} is a quite simple structure composed by  two masses connected by a massless rod. It is assume that this object is moving around a planet whose gravity field is approximated by the field of the attracting center. In general, the distance between the two points masses is considered to be much smaller that the distance between the satellite's center of mass and the attracting center of mass. Thus, it is common to neglect the influence of the attitude dynamics on the motion of the center of mass and treat it as an unperturbed Keplerian one.

\medskip

Rodnikov \cite{Rodnikov} studied equilibrium positions of a weight on a cable fixed to a
dumbbell--shaped space station moving along a circular geocentric orbit. This model
is composed by two masses coupled by a weightless rod, while the cable is
weightless and non-stretched. The equations of motion are stated when the motion
is produced in a single plane and the center of mass of the system moves along a
circular geocentric orbit. Moreover, the equilibrium configurations of the
system are obtained and the Lyapunov stability of configurations for two situations, first
when the station is composed of equal masses, second when masses at the ends of the station are different are analyzed.

\medskip

For the ``dumbbells--load'' system with two unilateral connections, all relative equilibria on the circular Keplerian orbit were established by Munitsina \cite{Munitsina}. Recall that a relative equilibria of the system is a point of the phase space giving an evolution which is  a one--parameter orbit of the action of the symmetry group of the system. These results were interpreted for studying the relative equilibria for which both connections are stretched in geometrical terms. The necessary and sufficient conditions for stability of the relative equilibria were stated.

\medskip

Celletti and Sidorenko \cite{Celletti} investigated the dumbbell satellite's attitude dynamics, when the center of mass moves on a Keplerian trajectory. They found a stable relative equilibrium position in the case of circular orbits which disappears as far as elliptic trajectories are considered. They replaced the equilibrium position by planar periodic motions and they proved this motion is unstable with respect to out-of-plane perturbations. They also gave some numerical evidences of the existence of stable spatial periodic motions.

\medskip

Burov et al. \cite{Burov2} considered the motion of a dumbbell--shaped body in an attractive Newtonian central field. They used the Poincare's theory to determine the conditions for the existence of families of system periodic motions depending on the arising small parameter and passing into some stable radial steady--state motion of the unperturbed problem as the small parameter tends to zero. They also proved that, each of the radial relative equilibria generates one family of such periodic motions, for sufficiently small parameter values. Furthermore, they studied the stability of the obtained periodic solutions in the linear approximation as well as these solutions were calculated up to terms of the first order in the small parameter.

\medskip

Guirao et al. \cite{Guirao} gave sufficient conditions for the existence of periodic solutions of the perturbed attitude dynamics of a rigid dumbbell satellite in a circular orbit.

\medskip

The statement of our main results is the following.

\begin{theorem}\label{th1}
Consider a dumbbell satellite moving in a gravity field generated by an oblate body considering the effect of the zonal harmonic parameter $A$. The pass trajectory of the mass center of the system is periodic and different from the classical one. If $A$ is equal to zero our solution coincides with the elliptical classical one.
\end{theorem}

Finally, considering the motion in the satellite approximation we complete the classical theory, stating the following result.

\begin{theorem}\label{th2}
The equations of motion in the satellite approximation can be reduced to Beletsky's equation when $A$ is equal to zero.
\end{theorem}

Note that Theorem \ref{th1} generalizes Celletti and Sidorenko \cite{Celletti}, Burov and Dugain \cite{Burov1} and Nakanishi et al. \cite{Nakanishi} due to oblateness parameter.

The structure of the paper is as follows. In Section \ref{sec2} we present the model description, the potential, the kinetic energy and the Lagrangian function of the system. In Section \ref{sec3} we present the morphology of the equations of motion and the equation of the mass center of the system. In Sections \ref{sec4} and \ref{sec5} we respectively provide proof of Theorems \ref{th1} and \ref{th2}. We remark that when $J_{2}=0$ is clear that the dynamics occurs on a plane, however when the coefficient $J_2$ is considered the effects of the gravitational potential are not the same for planes with different inclinations and a natural question is if there is an invariant plane for the dynamics. The answer of this fact is positive and it will be a key point in the proofs of our main result. In the Appendix  we provide a proof of this property.

\section{Model description}\label{sec2}

\subsection{Hypothesis}
We assume that the dumbbell satellite is formed by massless rod of length $l$ with to masses $m_{1}$ and $m_{2}$ placed at its ends. Let  consider $c$ the center of mass of the two masses moving in a gravity field generated by an oblate body whose mass $m$ having mass center located at $0$ where the distance between $0$ and $c$ is $r$ and $r\gg l$.

Let us consider the orbital reference frame $c$xy with origin at the dumbbell's center, and the polar coordinates of the center are $(r,\theta)$. While the rotation of the satellite relative to ray $oc$ will be determined by an angle $\Theta$. Furthermore we denote the reduced mass by $\mu$ and the sum of the two masses by $m_{s}$ where $\mu=m_{1}m_{2}/m_{s}$ and $m_{s}=m_{1}+m_{2}$, see Figure 1 for details.

\begin{center}
\begin{figure}
\includegraphics[width=80.85mm,height=70.01mm]{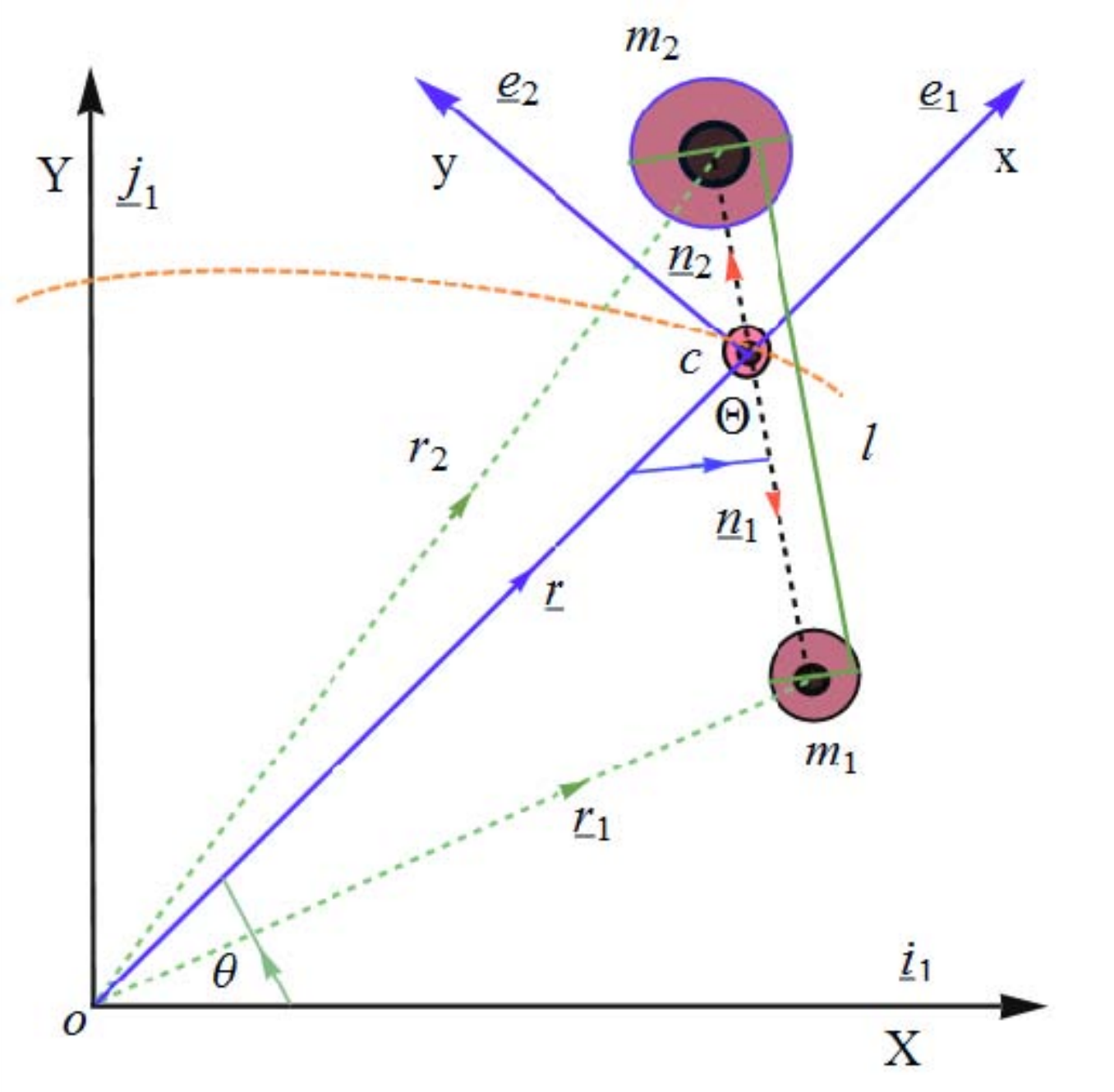}
\caption{The dumbbell satellite model}
\end{figure}
\end{center}

Now, we assume that $\underline{r}_{i}$ is the position vector of $m_{i}$ with
respect to $0$. Moreover, let the vector $\underline{n}_{i}$ denotes the position vector of $m_{i}$ with respect to
the center of mass of the dumbbell satellite, $i\in\{1,2\}$ .

Therefore, the magnitudes of the position vectors $\underline{r}_{i}$ are controlled by

\begin{equation}\label{1.1}
 r_{i}^{2}=r^{2}+n_{i}^{2}+2(-1)^{2-i}n_{i}r\cos\Theta
\end{equation}

where

\begin{equation}\label{1.2}
n_{i}=m_{3-i}l/m_{s}.
\end{equation}

\subsection{The potential of the model}
From the potential theory, the gravitational potential (any object has axial
symmetry $m_{0}$) experienced by the satellite $m$ will be controlled by  (see Murray and Dermott \cite{Murray})

\begin{equation}\label{ecunueva}
V=-\dfrac{Gm_{0}m}{r_{0}}\left[1-\sum_{n=2}^{+\infty}J_{n}\left(\dfrac{R}{r_{0}}\right)p_{n}(\sin(\delta))\right]
\end{equation}

where:

\begin{enumerate}
\item $G$ is the universal constant; $m_0$ is the mass of the oblate object and $m$ is the mass of the satellite;

\item $R$ is the mean radius of the oblate object;

\item $J_{n}$ is a dimensionless coefficient that characterizes the size of non--spherical components of the potential;

\item $r_{0}$ is the distance between $m_{0}$ and $m$;

\item $p_{n}\sin(\delta)$ are the Legendre polynomials of degree $n$;

\item $\delta$ denotes the latitude of the satellite.
\end{enumerate}

If the two bodies move in the same plane, then $\delta=0$ and equation \eqref{ecunueva} can be written as:

\begin{equation}\label{ecunueva1}
V=-\dfrac{Gm_{0}m}{r_{0}}\left[1-\sum_{n=2}^{+\infty}J_{n}\left(\dfrac{R}{r_{0}}\right)p_{n}(0)\right],
\end{equation}

where

$p_{2n}(0)=\dfrac{(-1)^{n} 2n!}{2^{2n}(n!)^2}$, $p_{2n+1}(0)=0$.

In the present model we shall consider the planar motion, for more details on it see the Appendix, and take the effect of the zonal harmonic up to $J_2$, hence equation \eqref{ecunueva1} can be rewritten as

\begin{equation*}\label{2}
V_{0}=-Gm_{0}m(\displaystyle \frac{1}{r_{0}}+\frac{J_{2}R^{2}}{2r_{0}^{3}}),
\end{equation*}

see \cite{M0,M1,M2,M3,M4,M5,M6,M7} for more details.

If we assume that $R$ represent the unit of distance, $m_{0}$ is also the unit
mass and denote $J_{2}$ by $A$. We have that the potential experienced by the masses $m_{1}$ and $m_{2}$ are $V_{1}$ and $V_{2}$ such that

\begin{equation}\label{3.1}
V_{1}=-Gm_{1}(\displaystyle \frac{1}{r_{1}}+\frac{A}{2r_{1}^{3}}),
\end{equation}

\begin{equation}\label{3.2}
V_{2}=-Gm_{2}(\displaystyle \frac{1}{r_{2}}+\frac{A}{2r_{2}^{3}}).
\end{equation}

Therefore the total potential $V$ can be written as

\begin{equation}\label{4}
V=-k(\displaystyle \frac{m_{1}}{r_{1}}+\frac{m_{2}}{r_{2}}+A(\frac{m_{1}}{2r_{1}^{3}}+\frac{m_{2}}{2r_{2}^{3}})),
 \end{equation}
where $k=G$ denotes the gravity parameter associated to the oblate body.

\subsection{The kinetic energy of the model}
Let the vectors $\underline{e}_{1}$ and $\underline{e}_{2}$ be an
orthogonal set of unitary vectors with $\underline{e}_{1}$ corresponding to the direction from $0$ to $c$.

Consider $\underline{i}$ and $\underline{j}$ be another orthogonal set of unitary vectors such that $\underline{i}$ is a vector in the direction of $x$ axis. Consequently the vectors of the locations $\underline{r}_{i}$ and associates velocities $\underline{v}_{i}$ of masses $m_{i}$ can be written as

$$\underline{r}_{i}=\underline{r}+\underline{n}_{i},$$

$$\displaystyle \underline{v}_{i}=\frac{d\underline{r}_{i}}{dt},$$

where

$$\underline{r}=r(\cos\theta\underline{i}+\sin\theta\underline{j}),$$

$$\underline{n}_{i}=(-1)^{i}n_{i}(\cos\Theta\underline{e}_{1}+\sin\Theta\underline{e}_{2}),$$

$$\underline{e}_{i}=(-1)^{i}\left(\cos \left(\theta + \dfrac{\pi}{i}\right) \underline{i}+
\sin\left(\theta + \dfrac{\pi}{i} \right) \underline{j}\right).$$

Therefore, after some calculations, we obtain

\begin{equation}\label{5}
v_{i}^{2}=\left\{\begin{array}{l}
\dot{r}^{2}+r^{2}\dot{\theta}^{2}+n_{i}^{2}(\dot{\theta}+\dot{\Theta})^{2}\\
-2(-1)^{i}\dot{r}n_{i}(\dot{\theta}+\dot{\Theta})\mathrm{s}\mathrm{i}\mathrm{n}\Theta\\
+2(-1)^{i}rn_{i}\dot{\theta}(\dot{\theta}+\dot{\Theta})\mathrm{c}\mathrm{o}\mathrm{s}\Theta
\end{array}\right\}.
\end{equation}

Since the kinetic energy of the dumbbell satellite system is

\begin{equation}\label{7}
T=\displaystyle \frac{1}{2}\sum_{i=1}^{2}m_{i}v_{i}^{2}.
\end{equation}

Substituting equation \eqref{5} into \eqref{7}, the kinetic energy can be written in the form

\begin{equation*}\label{8.1}
T=T_{s}+T_{r}.
\end{equation*}
where

\begin{equation}\label{8.2}
T_{s}=\displaystyle \frac{1}{2}m_{s}(\dot{r}^{2}+r^{2}\dot{\theta}^{2}).
\end{equation}

\begin{equation*}\label{8.3}
T_{r}=\displaystyle \frac{1}{2}\mu l^{2}(\dot{\theta}+\dot{\Theta})^{2}.
\end{equation*}

Hence

\begin{equation}\label{9}
T=\displaystyle \frac{1}{2}m_{s}(\dot{r}^{2}+r^{2}\dot{\theta}^{2})+\frac{1}{2}\mu l^{2}(\dot{\theta}+\dot{\Theta})^{2}.
\end{equation}

\subsection{The Lagrangian function of the model}
Since the Lagrange's function is defined by $L=T-V$ from equations \eqref{4} and \eqref{9} we get
\begin{equation}\label{10}
L=\displaystyle \frac{1}{2}m_{s}(\dot{r}^{2}+r^{2}\dot{\theta}^{2})+\frac{1}{2}\mu l^{2}(\dot{\theta}+\dot{\Theta})^{2}+k(\frac{m_{1}}{r_{1}}+\frac{m_{2}}{r_{2}}+A(\frac{m_{1}}{2r_{1}^{3}}+\frac{m_{2}}{2r_{2}^{3}})).
\end{equation}
Therefore, the equations of motion will be governed by
\begin{equation}\label{11}
\displaystyle \frac{d}{dt}(\frac{\partial L}{\partial\dot{\chi}})-\frac{\partial L}{\partial\chi}=0,\ \chi\in\{r,\theta,\Theta\}.
\end{equation}

\section{Equation of motion}\label{sec3}
\subsection{Equations of motion for the general case}
Substituting equation \eqref{10} into \eqref{11} when $\chi\in\{r,\theta,\Theta\}$ the equations of motion can
be written in the following form

\begin{equation}\label{12.2}
\begin{array}[c]{l}
(m_{s}r^{2}+\mu l^{2})\dot{\theta}+\mu l^{2}\dot{\Theta}=p_{\theta}=Q,\\

m_{s}(\displaystyle \ddot{r}-r(\frac{p_{\theta}-\mu l^{2}\dot{\Theta}}{m_{s}r^{2}+\mu l^{2}})^{2})=
-k\text{ }\left\{\begin{array}{l}
(\frac{m_{1}(r-n\mathrm{c}\mathrm{o}\mathrm{s}\Theta)}{r_{1}^{3}}+\frac{m_{2}(r+(l-n)\mathrm{c}\mathrm{o}\mathrm{s}\Theta)}{r_{2}^{3}})\\
+\frac{3}{2}A(\frac{m_{1}(r-n\mathrm{c}\mathrm{o}\mathrm{s}\Theta)}{r_{1}^{5}}+\frac{m_{2}(r+(l-n)\mathrm{c}\mathrm{o}\mathrm{s}\Theta)}{r_{2}^{5}})
\end{array}\right\},\\
\displaystyle \ddot{\Theta}+\frac{2\dot{r}(\mu l^{2}\dot{\Theta}-p_{\theta})}{r(m_{s}r^{2}+\mu l^{2})}=
\displaystyle -\frac{k(m_{s}r^{2}+\mu l^{2})\sin\Theta}{m_{s}lr} ( (\displaystyle \frac{1}{r_{1}^{3}}-\frac{1}{r_{2}^{3}})+\frac{3}{2} A (\displaystyle \frac{1}{r_{1}^{5}}-\frac{1}{r_{2}^{5}})),
\end{array}
\end{equation}
where $Q$ is constant and $n_{1}=n$, $n_{2}=l-n$ while $p_{\theta}$ is a constant expresses the angular momentum
conservation.

\subsection{Dumbbell's center of motion}

Since $(r,\theta)$ is the coordinate of the dumbbell's center, therefore the kinetic
energy $T_{s}$ is given by equation \eqref{8.2}, while the potential of the center of mass
$V_{s}$ is given by

\begin{equation*}\label{13}
V_{s}=-G m_{s}(\displaystyle \frac{1}{r}+\frac{A}{2r^{3}}).
\end{equation*}

Consequently, the Lagrange function $L_{s}$ of the center of mass can be represented
in the form
\begin{equation}\label{14}
L_{s}=\displaystyle \frac{1}{2}m_{s}(\dot{r}^{2}+r^{2}\dot{\theta}^{2})+G m_{s}(\frac{1}{r}+\frac{A}{2r^{3}}).
\end{equation}

Substituting equation \eqref{14} into \eqref{11} with $L=L_{s}$ and $\chi\in\{r,\theta\}$ and taking account that the equations of motion can be written on the form

$$\displaystyle \frac{d}{dt}(\frac{\partial L_{s}}{\partial\dot{r}})-\frac{\partial L_{s}}{\partial r}=0,$$

$$\displaystyle \frac{d}{dt}(\frac{\partial L_{s}}{\partial\dot{\theta}})-\frac{\partial L_{s}}{\partial\theta}=0,$$

we state that the motion of dumbbell's center will be controlled by

\begin{equation}\label{15}
\begin{array}[c]{l}
\displaystyle \ddot{r}-r\dot{\theta}^{2}=-k(\frac{1}{r^{2}}+\frac{3A}{2r^{4}}),\\
m_{s}r^{2}\dot{\theta}=F$ or $r^{2}\dot{\theta}=h,\\
\dfrac{1}{2}\dot{r}^{2}-\dfrac{k}{r}-\dfrac{6kA}{r^{3}}=E,
\end{array}
\end{equation}
where $F$ is a constant, $h$ is the angular momentum which is constant too, that can be evaluated by the initial
conditions and $E$ is the preservation of the total energy for the dumbbell's center.

\medskip

Let be $r=\displaystyle \frac{1}{u}$ , consequently

\begin{equation}\label{16}
\displaystyle \frac{d^{2}u}{d\theta^{2}}+u=\frac{k}{h^{2}}(1+\frac{3}{2}Au^{2}).
\end{equation}

\begin{equation}\label{new}
\begin{array}[c]{l}
\displaystyle \dot{\theta}=-h u^{2},\\
\dfrac{1}{2}h^{2}\left(\dfrac{du}{d\theta}\right)^{2}-k u-6kAu^{3}=E.
\end{array}
\end{equation}
It is worth mentioning that the system of equations \eqref{16} does not represent only dumbbell's center motion, it represents too
the motion of two--body problem under the effect of the zonal harmonic motion which can be reduced to the motion of the
classical case when $A=0$. Therefore our results on the dumbbell's center of motion can be applied it to the motion of
two--body problem.

Now let us go back to dumbbell's center motion in which we can be assumed that this motion follows a Kepler's type
orbit when the effect of oblateness parameter is absent ($A=0$). Consequently the solution can be written as
$$r_{0}=\displaystyle \frac{h^{2}/k}{(1+e_{0}\cos\theta)}$$
where
$r_{0}=\displaystyle \frac{1}{u_{0}}$ , $e_{0}$ is the orbit eccentricity such that $0\leq e_{0}<1$, in the framework of
elliptic orbits and $\theta$ is a true anomaly of the center of mass.
When $\theta=0,\ u_{0}=\displaystyle \frac{1}{r_{p}}=\frac{k}{h^{2}}(1+e_{0}),\ r_{p}=a(1-e_{0})$ is the pericenter (periapsis)
and $a$ is a semi--major axis, see Figures 2 and 3.

\medskip

Now we look for solutions in the form $u(\theta,\epsilon)$ under the condition $ 0<\epsilon \ll 1$. Since $A=J_{2}$  and $J_{2}\in[1\times 10^{-3},1\times 10^{-6}]$ for the most of celestial bodies then we can replace $\epsilon$ by A. In addition, this solution must hold the initial conditions
$$u(0,\displaystyle A)=\frac{1}{r_{p}},$$
$$D_{\theta}u(0,A)=0.$$

Therefore, we search for straight forward expansion of an asymptotes solution as a tends to zero in the following form
\begin{equation}\label{20}
u(\theta,A)=u_{0}(\theta)+Au_{1}(\theta)+o(A^{2}).
\end{equation}
\medskip

The effect of the zonal harmonic of $J_{2}$ is taking account, but the perturbation due to $J_{2}$ is of order about $10^{-3}$ of the unperturbed main term $(m_{1}/r_{1})$ or $(m_{2}/r_{2})$. While all other coefficients of zonal harmonic are about $10^{-6}$ or less. Therefore, it is sufficient from practical point of view, we take the expansion in equation \eqref{20} up to $A$. On the other hand, $o(A^{2})$ represents the effect of the zonal harmonic $J_{4}$ while our potential does not contain the zonal harmonic $J_{4}$. Consequently we truncate the expansion in equation \eqref{20} up to the linear term $A$. In this case the leading--order perturbation equations are
$$\displaystyle D_{\theta\theta}u_{0}+u_{0}=\frac{k}{h^{2}},$$
$$\displaystyle D_{\theta\theta}u_{1}+u_{1}=\frac{3}{2}\frac{k}{h^{2}}u_{0}^{2}.$$

Under the conditions $u_{0}(0)=\displaystyle \frac{1}{r_{p}}$ , $D_{\theta}u_{0}=0,\ u_{1}(0)=0$ and $D_{\theta}u_{1}=0$, hence the
solution is governed by

$$u_{0}=\displaystyle \frac{k}{h^{2}}(1+e_{0}\cos\theta)$$
and
$$u_{1}=\displaystyle \frac{3k^{3}}{2h^{6}}[1+\frac{1}{2}e_{0}^{2}-(1+\frac{1}{3}e_{0}^{2})\cos\theta+e_{0}\theta\sin\theta-\frac{1}{6}e_{0}^{2}\cos 2\theta].$$
Therefore, the general expression of the dumbbell's center motion up to $o(A)$ will be
governed by
\begin{equation}\label{22}
u(\theta,A)=u_{0}(\theta)+Au_{1}(\theta).
\end{equation}

\section{Proof of Theorem \ref{th1}}\label{sec4}

Since equation \eqref{22} represents a solution which contains a secular term that grows in $\theta$. As a result, the expansion is not uniformly valid in $\theta$ and breaks down when $\theta=o(A)$, furthermore $A u_{1}$ is no longer a small correction of $u_{0}$. But convergent series approximation of the periodic solution can be determined by the continuation method known as the Lindstedt--Poincare's technique.

\medskip

Since equation \eqref{16} is a second order differential equation, it describes a dynamical system in which A is a small parameter. Consequently if $A=0$ the system will be reduced to a harmonic oscillator which has a solution with period $ T=2\pi /\omega_{0}$ where $\omega_{0}=1$.

\medskip

The continuation method enables us to construct a periodic solution for $A\neq 0$. If we consider that the angular velocity changes due to the non--linear terms, the asymptotic solution $u(\theta,\ A)$ and the angular velocity $\omega$ of the dynamical system can be expanded as
\begin{equation}\label{23}
\begin{array}[c]{l}
u(\theta,A)=u_{0}(\theta)+Au_{1}(\theta)+A^{2}u_{2}(\theta)+\ldots\\
\omega =1+A \omega_{1}+A^{2}\omega_{2}+\ldots
\end{array}
\end{equation}

To construct a uniformly valid solution, we will introduce a stretched variable $\tau=\omega\theta$, therefore
\begin{equation}\label{24}
\begin{array}[c]{l}
\displaystyle \frac{d}{d\theta}=\omega\frac{d}{d\tau},\\
\displaystyle \frac{d^{2}}{d\theta^{2}}=\omega^{2}\frac{d^{2}}{d\tau^{2}}.
\end{array}
\end{equation}

Substituting equations \eqref{24} into \eqref{16} we obtain
\begin{equation}\label{25}
\omega^{2}\displaystyle \frac{d^{2}u}{d\tau^{2}}+u=\frac{k}{h^{2}}(1+\frac{3}{2}Au^{2}).
\end{equation}

Now, under the following conditions

$$u(\displaystyle 0,A)=\frac{1}{r_{p}},$$
$$u_{\tau}(0,A)=0,$$
$$u(\tau+2\pi,A)=u(\tau,A),$$
we insert the series expansion \eqref{23} into \eqref{25} and equating terms of the same order in $A$ with keeping the terms up to first order of $A$, we obtain the following:

\begin{itemize}
\item The coefficient of $A^{0}$ gives a homogeneous equation in the form
$$
\frac{d^{2}u_{0}}{d\tau^{2}}+u_{0}=\frac{k}{h^{2}}
$$

where$\text{ }u_{0}(0)=\frac{1}{r_{p}}\text{ , }\frac{du_{0}(0)}{d\tau}=0\;\text{and}\text{ }u_{0}(\tau+2\pi,A)=u_{0}(\tau,A)
$
with a solution
\begin{equation}\label{28}
u_{0g}(\tau)=\frac{k}{h^{2}}(1+e_{0}\cos\tau),
\end{equation}

being

\begin{equation}\label{erre}
r=\dfrac{h^{2}/k}{1+e_{0}\cos\tau}.
\end{equation}

\item The coefficient of A gives a non--homogeneous equation in the form
\begin{equation}\label{29}
\displaystyle \frac{d^{2}u_{1}}{d\tau^{2}}+u_{1}=a_{1}+a_{2}\cos\tau+a_{3}\cos 2\tau
\end{equation}

where $u_{1}(0)=0$ , $\displaystyle \frac{du_{1}(0)}{d\tau}=0$ and

$$a_{1}=\displaystyle \frac{3k^{3}}{2h^{6}}(1+\frac{1}{2}e_{0}^{2}),$$
$$a_{2}=\displaystyle \frac{3e_{0}k^{3}}{h^{6}}(1+\frac{2\omega_{1}h^{4}}{3k^{2}}),$$
$$a_{3}=\displaystyle \frac{3e_{0}^{2}k^{3}}{4h^{6}}$$
with a particular solution
\begin{equation*}
u_{1}(\tau+2\pi,A)=u_{0}(\tau,A)
\end{equation*}

$$u_{1p}=a_{1}+\displaystyle \frac{1}{2}a_{2}\cos\tau+\frac{1}{2}a_{2}\tau\sin\tau-\frac{1}{3}a_{3}\cos 2\tau.$$
This solution contain a secular term $a_{2}\tau\sin\tau/2$, to avoid this term and the
solution becomes periodic we have to equate it coefficient by zero, hence
$$\omega_{1}=-\displaystyle \frac{3k^{2}}{2h^{4}}.$$
\end{itemize}

Therefore the general solution of equation \eqref{29} is controlled by
\begin{equation}\label{33}
u_{1g}=\displaystyle \frac{3k^{3}}{2h^{6}}(1+\frac{1}{2}e_{0}^{2})-\frac{k^{3}}{2h^{6}}(3+e_{0}^{2})\cos\tau-\frac{e_{0}^{2}k^{3}}{4h^{6}}\cos 2\tau.
\end{equation}

Substituting equations \eqref{28} and \eqref{33} into \eqref{23}, the general solution of equation \eqref{25} becomes

$$u=\displaystyle \frac{k}{h^{2}}(1+k_{1})[1+(\frac{e_{0}-k_{2}}{1+k_{1}})\cos\tau+\frac{k_{3}}{1+k_{1}}\cos 2\tau]$$

where
$$
k_{1}=\dfrac{3Ak^{2}}{4h^{4}}(2+e_{0}^{2}),
$$
$$
k_{2}=\dfrac{Ak^{2}}{2h^{4}}(3+e_{0}^{2}),
$$
$$
k_{3}=-\frac{Ak^{2}e_{0}^{2}}{4h^{4}},
$$
$$
\tau=(1-\frac{3Ak^{2}}{2h^{4}})\theta.
$$
Therefore
\begin{equation}\label{36}
r=\frac{h^{2}/\overline{k}}{(1+e\cos\tau+\overline{e}\cos 2\tau)},
\end{equation}
with
$$
\overline{k}=k(1+k_{1}),
$$
$$
e=(\frac{e_{0}-k_{2}}{1+k_{1}}),
$$
$$
\overline{e}=\dfrac{k_{3}}{1+k_{1}}.
$$
In short, it is clear that the trajectory of the mass center differs from that assumed by
Celletti and Sidorenko \cite{Celletti}, Burov and Dugain \cite{Burov1} and Nakanishi et al. \cite{Nakanishi}
due to oblateness parameter. Although, this solution is periodic. While this
trajectory is the same as their solutions when the effect of oblateness is ignored.
Since $e_{0}<1$ and $A<<1$ as a result $Ae_{0}^{2}<<1$ is very small. Therefore, if we
neglect all terms that include $Ae_{0}^{2}$, the equation \eqref{36} will be reduced to

\begin{equation}\label{erre}
r=\displaystyle \frac{h^{2}/\overline{k}}{(1+e\cos\tau)},
\end{equation}

$$\displaystyle \overline{k}=k(1+\dfrac{3Ak^{2}}{2h^4}),$$

$$e=e_{0}-\displaystyle \dfrac{3Ak^{2}}{2h^4}(1+e_{0}).$$
This means that the trajectory of the mass center is elliptic as the classical case
with the decreasing of the elliptical parameter and the eccentricity, ending the proof. \bbox

\begin{remark}
Taking account the oblateness effect we have proved that the solution is periodic, see equation \eqref{36}. While for the small
value of the parameter A, we have elliptical solutions as in the classical case with the decreasing in the elliptical parameter, see \eqref{erre}.

Thus, there is no discontinuity and the solution varies smoothly as $A$ approaches zero.
\end{remark}

\begin{center}
\begin{figure}
\includegraphics[width=90.85mm,height=80.01mm]{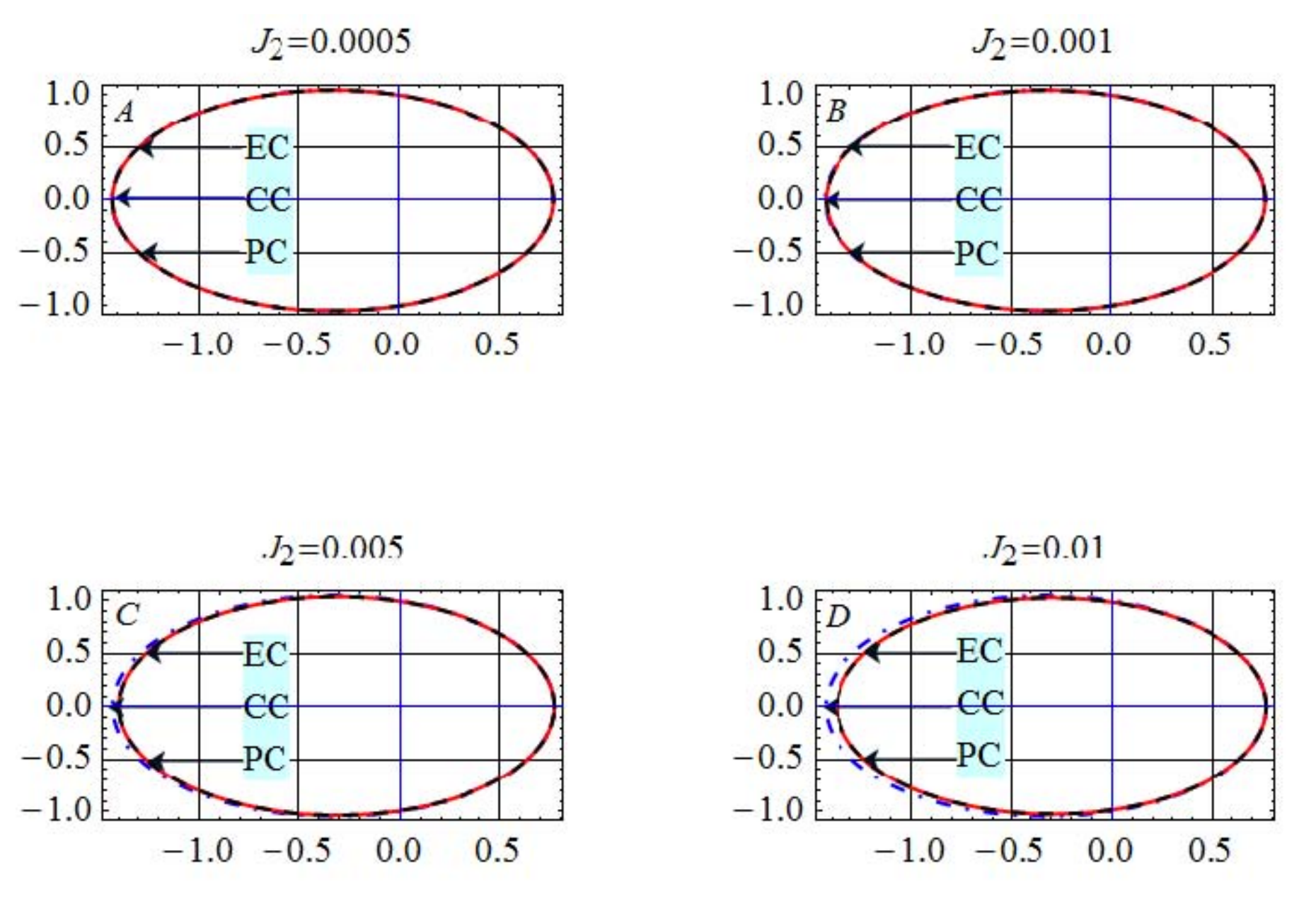}
\caption{Variation in the trajectory of dumbbell's center when $e_{0}=0.3$ for different values of zonal harmonic parameter.}
\end{figure}
\end{center}

\begin{center}
\begin{figure}
\includegraphics[width=90.85mm,height=80.01mm]{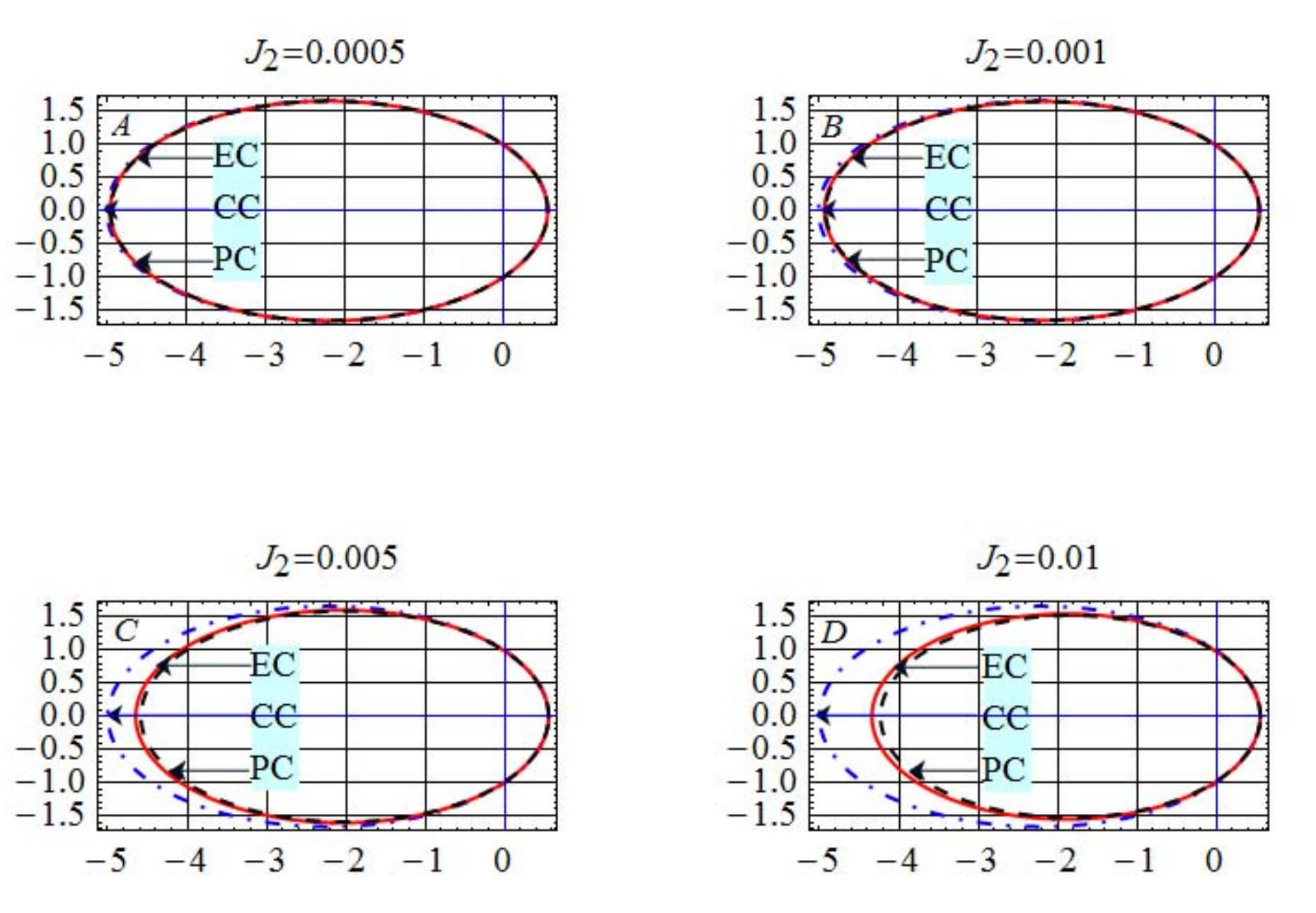}
\caption{Variation in the trajectory of dumbbell's center when $e_{0}=0.8$ for different values of zonal harmonic parameter.}
\end{figure}
\end{center}

\begin{note}
Figures 2 and 3 represent the changes in the trajectory of the dumbbell's center corresponding to the
changes in the eccentricity and in the zonal harmonic parameter, here we have considered that $k$ and $h$ are equal to 1. We denote the curves representing the classical case
(the effect of zonal harmonic is switched off) by (CC). If the effect of zonal harmonic is consider ignoring all
terms with coefficients $Ae_{0}^2$ the curves will be denoted by (EC). Finally, by (PC) we denote the general trajectory.

\smallskip

We observe that all trajectories are quasi--elliptical and
the decreasing in the ellipse parameters is very small for small values of classical eccentricity $e_{0}$ and the zonal
harmonic parameter $J_2$, see some cases of Figure 2. While for some relative large values of the
classical eccentricity $e_{0}$ the decreasing in the ellipse parameter is observed especially when the parameter of zonal
harmonic is assigned by big value.
\end{note}

\section{Proof of Theorem \ref{th2}}\label{sec5}

We shall start by the deduction of the equations of motion in satellite approximation. Indeed, substituting equations
\eqref{1.1} and \eqref{1.2} into \eqref{4}, the approximation of the potential energy can be written as

\begin{equation}\label{nuevaecumoro}
\displaystyle V=-k(m_{s}(\frac{1}{r}+\frac{A}{2r^{3}})+\frac{\mu l^{2}}{2r^{3}}(3\cos^{2}\Theta-1)).
\end{equation}

In this potential we neglect all terms that contain coefficients $(1/r)$ with power four or more, since $l \ll r$. Therefore the Lagrangian function becomes

\begin{equation}\label{40}
\begin{array}[c]{l}
L=\displaystyle \frac{1}{2}m_{s}(\dot{r}^{2}+r^{2}\dot{\theta}^{2})+\frac{1}{2}\mu l^{2}(\dot{\theta}+\dot{\Theta})^{2}\\
+k(m_{s}(\displaystyle \frac{1}{r}+\frac{A}{2r^{3}})+\frac{\mu l^{2}}{2r^{3}}(3\cos^{2}\Theta-1)).
\end{array}
\end{equation}

Substituting equation \eqref{40} into \eqref{11}, the approximation equations of motion are
\begin{equation}\label{41}
\begin{array}[c]{l}
m_{s}(\displaystyle \ddot{r}-r\dot{\theta}^{2})=-k(m_{s}(\frac{1}{r^{2}}+\frac{3A}{2r^{4}})+\frac{3\mu l^{2}}{2r^{4}}(3\cos^{2}\Theta-1)),\\
(m_{s}r^{2}+\mu l^{2})\dot{\theta}+\mu l^{2}\dot{\Theta}=p_{\theta},\\
\displaystyle \mu l^{2}(\ddot{\theta}+\ddot{\Theta})=-\frac{3k\mu l^{2}}{r^{3}}\cos\Theta\sin\Theta.
\end{array}
\end{equation}

Now replacing the independent variable $t$ with the starched variable $\tau$ where
$\tau=\omega\theta$ and $r^{2}\dot{\theta}=h$ therefore, it is possible to write $\dot{\tau}=\Omega(\tau)$ such that

\begin{equation}\label{42}
\displaystyle \Omega(\tau)=\frac{\omega\overline{k}^{2}}{h^{3}}(1+e\cos\tau+\overline{e}\cos 2\tau)^{2}.
\end{equation}

Hence

\begin{equation}\label{43}
\begin{array}[c]{l}
\displaystyle \frac{d}{dt}=\Omega\frac{d}{d\tau}\vspace{0.5cm},\\

\displaystyle \frac{d^{2}}{dt^{2}}=\Omega^{2}\frac{d^{2}}{d\tau^{2}}+\Omega\Omega^{\prime}\frac{d}{d\tau},
\end{array}
\end{equation}
where $(\cdot)^{\prime}$ means $\displaystyle \frac{d}{d\tau}$.

Inserting equations \eqref{43} into \eqref{41} and using equation \eqref{42}, we obtain

\begin{equation}\label{44}
(1+e\cos\tau+\overline{e}\cos 2\tau)\Theta^{\prime\prime}-2(e\displaystyle \sin\tau+2\overline{e}\sin 2\tau)(\frac{1}{\omega}+
\Theta^{\prime})
+\displaystyle \frac{3k}{\omega^{2}\overline{k}}\cos\Theta\sin\Theta=0.
\end{equation}

Since  $\tau=\omega\theta$, we can rewrite equation \eqref{44} in the form

\begin{equation}\label{45}
\begin{array}[c]{l}
(1+e\cos \omega\theta +\overline{e}\cos 2\omega\theta) \displaystyle \frac{d^{2}\Theta}{d\theta^{2}}
-2\omega(e\sin \omega\theta +2\overline{e}\sin 2 \omega\displaystyle \theta)(1+\frac{d\Theta}{d\theta})\\
+\displaystyle \frac{3k}{\overline{k}}\cos\Theta\sin\Theta=0
\end{array}
\end{equation}
where
$$\omega=1-\dfrac{3Ak^{2}}{2h^{4}},$$
$$\displaystyle \overline{e}=\frac{Ak^{2}e_{0}^{2}}{4h^{4}},$$
and
$$\displaystyle \overline{k}=k[1+\dfrac{3Ak^{2}}{4h^{4}}(2+e_{0}^{2})].$$

Now, for finishing only remark that equation \eqref{45} can be reduced to Beletsky's equation,  see \cite{Beletsky2} for more details, if we assume the oblateness effect is not consider (i.e., $A =0$) obtaining the relation

$$(1+e\displaystyle \cos\theta)\frac{d^{2}\Theta}{d\theta^{2}}-2e\sin\theta\frac{d\Theta}{d\theta}+3\cos\Theta\sin\Theta=2e\sin\theta,$$
which ends the proof. \bbox

 \section*{Appendix}

 Let us introduce now an inertial reference frame $\mathcal{I}(\mathbf{O};$
$\mathbf{E}_{1}$,$\mathbf{E}_{2}$,$\mathbf{E}_{3})$. The coordinates of a
generic vector in this reference system are denoted by $\mathbf{x=(}%
x,y,z)_{\mathcal{I}}$ . Recall that we are considering a dumbbell formed by two material
points $\mathbf{M}_{1}$, of mass $m_{1}$ and $\mathbf{M}_{2}$ of mass $m_{2}$
rigidly connected by a segment of constant length $l$ and negligible mass
mutually attracted by a gravitational potential due to nearly spherical body
$\mathbf{M}$. Recall that the potential is given by
\[
\mathcal{V}(\mathbf{x})=-\frac{GM}{\left\Vert \mathbf{x}\right\Vert }\left(
1-\sum_{n=2}^{\infty}J_{n}\left(  \frac{R}{\left\Vert \mathbf{x}\right\Vert
}\right)  ^{n}P_{n}\left(  \frac{z}{\left\Vert \mathbf{x}\right\Vert }\right)
\right)
\]
with $G$ is the gravitational constant, $M$ the mass of the body, and $R$ the
equatorial radius. $P_{n}(u)$ is the Legendre polynomial of degree $n$ and
argument $u$, and the $J_{n}$ are constant coefficients characterizing the
potential of the body $M.$ We can define a rotating frame $\mathcal{R}%
(\mathcal{G};\mathbf{e}_{1}$,$\mathbf{e}_{2}$,$\mathbf{e}_{3})$, with
$\mathcal{G}$ be the center of masses of the dumbbell, such that the unitary
vector $\mathbf{e}_{3}$ is directed along the dumbbell towards the point
$\mathbf{M}_{2}$ and $\mathbf{e}_{1}$, $\mathbf{e}_{2}$ are two orthonormal
vectors, perpendicular to $\mathbf{e}_{3}$. In this frame, the principal
moments of inertia $(I_{1},I_{2},I_{3})$ of the dumbbell are%
\[
I_{1}=I_{2}=\mu l^{2},\text{ }I_{3}=0
\]
with
\[
\mu=\frac{m_{1}m_{2}}{m_{s}}%
\]
where $m_{s}=m_{1}+m_{2}$
and
\[%
\begin{array}
[c]{cc}%
l_{1}=n=\dfrac{m_{2}l}{m_{s}},\text{ } & l_{2}=l-n=\dfrac{m_{1}l}{m_{s}}%
\end{array}
\]
the distances from $\mathbf{M}_{1}$ and $\mathbf{M}_{2}$ to $\mathcal{G}.$

The attitude of the dumbbell is given by two angles, namely nutation $\Theta$
and precession $\phi$. The coordinates of points $M_{1}$ and $M_{2}$ in the
space frame $\mathcal{S}$ are%
\[%
\begin{array}
[c]{l}%
\mathbf{M}_{1}\equiv-l_{1}\left(  \sin\Theta\sin\phi,-\sin\Theta\cos\phi
,\cos\phi\right)  _{\mathcal{S}}\medskip\\
\mathbf{M}_{2}\equiv l_{2}\left(  \sin\Theta\sin\phi,-\sin\Theta\cos\phi
,\cos\phi\right)  _{\mathcal{S}}.
\end{array}
\]

The coordinates of $\mathcal{G}$ respect to the inertial frame $\mathcal{I},$
using cylindrical coordinates are
\[
\mathcal{G}\equiv(r\cos\theta,r\sin\theta,z)_{\mathcal{I}}.
\]
Using the Koenig's Theorem, the Lagrangian of the dumbbell is $\mathcal{L}\left(  \mathbf{P};\mathbf{V}_{\mathbf{P}}\right)$ equal to
\[
\frac{m_{s}}%
{2}\left(  \left(  \frac{dr}{dt}\right)  ^{2}+r^{2}\left(  \frac{d\theta}%
{dt}\right)  ^{2}+\left(  \frac{dz}{dt}\right)  ^{2}\right)  +\frac{\mu l^{2}%
}{2}\left(  \left(  \frac{d\phi}{dt}\right)  ^{2}+\left(  \frac{d\Theta}%
{dt}\right)  ^{2}\sin^{2}\phi\right)  -\mathcal{U}(\mathbf{P})
\]

with $(\mathbf{P};\mathbf{V}_{P})=\left(  r,z,\theta,\Theta,\phi;\frac{dr}%
{dt},\frac{dz}{dt},\frac{d\theta}{dt},\frac{d\Theta}{dt},\frac{d\phi}%
{dt}\right),$  and
\[
\mathcal{U}(\mathbf{P})=\mathcal{V}(\mathbf{x}_{\mathbf{M}_{1}})+\mathcal{V}%
(\mathbf{x}_{\mathbf{M}_{2}}).
\]
The coordinates of $\mathbf{x}_{\mathbf{M}_{1}}$ and $\mathbf{x}%
_{\mathbf{M}_{2}}$ are%
\[%
\begin{array}
[c]{c}%
\mathbf{x}_{\mathbf{M}_{1}}\equiv(r\cos\theta-l_{1}\sin\Theta\sin\phi
,r\sin\theta+l_{1}\sin\Theta\cos\phi,z-l_{1}\cos\phi)_{\mathcal{I}}\medskip\\
\mathbf{x}_{\mathbf{M}_{2}}\equiv(r\cos\theta+l_{2}\sin\Theta\sin\phi
,r\sin\theta-l_{2}\sin\Theta\cos\phi,z+l_{2}\cos\phi)_{\mathcal{I}}%
\end{array}
\]
and
\[%
\begin{array}
[c]{l}%
\left\Vert \mathbf{x}_{\mathbf{M}_{1}}\right\Vert ^{2}=r^{2}+z^{2}+l_{1}%
^{2}-2l_{1}\left(  z\cos\phi+r\sin\phi\sin\left(  \Theta-\theta\right)
\right)  \medskip\\
\left\Vert \mathbf{x}_{\mathbf{M}_{2}}\right\Vert ^{2}=r^{2}+z^{2}+l_{2}%
^{2}+2l_{2}\left(  z\cos\phi+r\sin\phi\sin\left(  \Theta-\theta\right)
\right)
\end{array}
.
\]

The potential of the system has the following expression%
\[
\begin{array}
[c]{c}%
\mathcal{U}(r,z,\Theta-\theta,\phi)=-\left(  \dfrac{GMm_{1}}{\left\Vert
\mathbf{x}_{\mathbf{M}_{1}}\right\Vert }\left(  1-\sum_{n=2}^{\infty}%
J_{n}\left(  \frac{R}{\left\Vert \mathbf{x}_{\mathbf{M}_{1}}\right\Vert
}\right)  ^{n}P_{n}\left(  \frac{z-l_{1}\cos\phi}{\left\Vert \mathbf{x}%
_{\mathbf{M}_{1}}\right\Vert }\right)  \right)  +\right.  \medskip\text{ }\\
\left.  \dfrac{GMm_{2}}{\left\Vert \mathbf{x}_{\mathbf{M}_{2}}\right\Vert
}\left(  1-\sum_{n=2}^{\infty}J_{n}\left(  \frac{R}{\left\Vert \mathbf{x}%
_{\mathbf{M}_{2}}\right\Vert }\right)  ^{n}P_{n}\left(  \frac{z+l_{2}\cos\phi
}{\left\Vert \mathbf{x}_{\mathbf{M}_{2}}\right\Vert }\right)  \right)
\right)
\end{array}
\]

\subsection*{A.1 Hamiltonian expressions}

From the expressions of the kinetic energy and the potential, we can derive
the Hamiltonian%
$$
\mathcal{H(}\mathbf{P};T\text{ }\mathbf{V}_{P})=\frac{1}{2m_{s}}\left(
P_{r}^{2}+\frac{P_{\theta}^{2}}{r^{2}}+P_{z}^{2}\right)  +\frac{1}{2\mu l^{2}%
}\left(  \frac{P_{\Theta}^{2}}{\sin^{2}\phi}+P_{\phi}^{2}\right)
+\mathcal{U}(r,z,\Theta-\theta,\phi)
$$
with
\[
\mathcal{(}\mathbf{P};T\text{ }\mathbf{V}_{P})=\left(  r,z,\theta,\Theta
,\phi;P_{r},P_{\theta},P_{z},P_{\Theta},P_{\phi}\right)  .
\]

The angles $\theta$ and $\Theta$ appear only as the difference $\Theta-\theta
$, we can reduce the order of the Hamiltonian by means of the canonical
transformation
\[
(\,\theta,\Theta-\theta,P_{\theta},P_{\Theta})\rightarrow(\lambda,\psi
,P_{\psi}-P_{\lambda},P_{\psi}).
\]
The new Hamiltonian is
\begin{equation*}
(A.1)
\mathcal{H}=\frac{1}{2m_{s}}\left(  P_{r}^{2}+\frac{(P_{\psi}-P_{\lambda}%
)^{2}}{r^{2}}+P_{z}^{2}\right)  +\frac{1}{2\mu l^{2}}\left(  \frac{P_{\psi
}^{2}}{\sin^{2}\phi}+P_{\phi}^{2}\right)  +\mathcal{U}(r,z,\psi,\phi).
\label{Hamiltonian}%
\end{equation*}
The variable $\lambda$ is cyclic and the momentum $P_{\lambda}$ is a constant
of the motion. The Hamiltonian itself is another integral.

\subsection*{A.2 Equations of the motion}

The Hamiltonian equations of the motion are
\begin{equation*}
(A.2.1)
\begin{array}
[c]{cc}
\dfrac{dr}{dt}=\dfrac{P_{r}}{\mu},& \dfrac{dP_{r}}{dt}
=\dfrac{(P_{\psi}-P_{\lambda})^{2}}{\mu r^{3}}-\dfrac{\partial\mathcal{U}
}{\partial r}, \\

\dfrac{dz}{dt}=\dfrac{P_{z}}{\mu}, & \dfrac{dP_{z}}{dt}=-\dfrac{\partial\mathcal{U}}{\partial z},\\

\dfrac{d\psi}{dt}=\dfrac{P_{\psi}-P_{\lambda}}{\mu r^{2}}+\dfrac{P_{\psi}}{\mu
l^{2}\sin^{2}\phi}, & \dfrac{dP_{\psi}}{dt}=-\dfrac
{\partial\mathcal{U}}{\partial\psi}, \\
 \dfrac{d\phi}
{dt}=\dfrac{P_{\phi}}{\mu l^{2}}, & \dfrac{dP_{\phi}}
{dt}=\dfrac{P_{\psi}^{2}\cos\phi}{\mu l^{2}\sin^{3}\phi}-\dfrac{\partial
\mathcal{U}}{\partial\phi}.
\end{array}
\label{Equations}
\end{equation*}

\noindent \textbf{Theorem.}
\emph{The equations} (A.2.1) \emph{has an invariant manifold given by}
$z\equiv0,P_{z}\equiv0,\phi\equiv\pi/2$ and $P_{\phi}\equiv0.$

\noindent \emph{Proof.}
Using the equations
\[%
\begin{array}
[c]{ll}%
\dfrac{dz}{dt}=\dfrac{P_{z}}{\mu},\text{ }\medskip & \dfrac{dP_{z}}%
{dt}=-\dfrac{\partial\mathcal{U}}{\partial z},\medskip\\
\dfrac{d\phi}{dt}=\dfrac{P_{\phi}}{\mu l^{2}},\text{ }\medskip &
\dfrac{dP_{\phi}}{dt}=\dfrac{P_{\psi}^{2}\cos\phi}{\mu l^{2}\sin^{3}\phi
}-\dfrac{\partial\mathcal{U}}{\partial\phi}.
\end{array}
\]
the result is immediate.
\bbox

The Hamiltonian (A.1) restricted to the invariant manifold is%
\[
\mathcal{H}=\frac{1}{2m_{s}}\left(  P_{r}^{2}+\frac{P_{\psi}^{2}}{l^{2}}%
+\frac{(P_{\psi}-P_{\lambda})^{2}}{r^{2}}\right)  +\mathcal{U}_{1}(r,\psi).
\]
with%
\[
(A.2.2)
\begin{array}
[c]{c}%
\mathcal{U}_{1}(r,\psi)=-GM\left[ \left(  \dfrac{m_{1}}{\sqrt{r^{2}+l_{1}%
^{2}-2l_{1}r\sin\psi}}+\dfrac{m_{2}}{\sqrt{r^{2}+l_{2}^{2}+2l_{2}r\sin\psi}%
}\right)  +\right.  \medskip\\
\left.  R^{2}J_{2}\left(  \dfrac{m_{1}}{\left(  \sqrt{r^{2}+l_{1}^{2}%
-2l_{1}r\sin\psi}\right)  ^{3}}+\dfrac{m_{2}}{\left(  \sqrt{r^{2}+l_{2}%
^{2}+2l_{2}r\sin\psi}\right)  ^{3}}\right)  +O(J_{4})\right]
\end{array}
\]
If $r>>l$ and $R=1$, $M=1$, $k=G$,  $J_{2}=A$ we obtain
\[
(A.2.3)\;\;
\mathcal{U}_{1}(r,\psi)=-k\left( m_{s}  \left(  \frac
{1}{r}+\frac{A}{2r^{3}}\right)  +\frac{\mu l^{2}}{2r^{3}}\left(  3\cos
^{2}\psi-1\right)  \right)  .
\]

\subsection*{A.3 The Lagrangian function.}

The Lagrangian are%
\[
\mathcal{L}\left(  r,\lambda,\psi,\frac{dr}{dt},\frac{d\lambda}{dt}%
,\frac{d\psi}{dt}\right)  =\frac{m_{s}}{2}\left(  \left(  \frac{dr}%
{dt}\right)  ^{2}+r^{2}\left(  \frac{d\lambda}{dt}\right)  ^{2}\right)
+\frac{\mu l^{2}}{2}\left(  \frac{d\left(  \lambda+\psi\right)  }{dt}\right)
^{2}-\mathcal{U}_{1}(r,\psi)
\]
and the second order equations of the motion are given by%
\[
(A.3)
\begin{array}
[c]{l}%
m_{s}\left(  \dfrac{d^{2}r}{dt^{2}}-r\left(  \dfrac{d\lambda}{dt}\right)
^{2}\right)  =-\dfrac{\partial \mathcal{U}_{1}}{\partial r},\medskip\\
\mu l^{2}\left(  \dfrac{d^{2}\lambda}{dt^{2}}+\dfrac{d^{2}\psi}{dt^{2}%
}\right)  =-\dfrac{\partial \mathcal{U}_{1}}{\partial\psi},\medskip\\
m_{s}r^{2}\dfrac{d\lambda}{dt}+\mu l^{2}\left(  \dfrac{d\lambda}{dt}%
+\dfrac{d\psi}{dt}\right)  =\text{constant}.
\end{array}
\]

It is clear that if we replace the symbols $\lambda$ by $\theta$ and $\psi$ by $\Theta$ the equation (A.2.3) is the same of equation \eqref{nuevaecumoro}.
Also the system of equations (A.3) becomes into the system of equations \eqref{12.2} when $\mathcal{U}_{1}$ is represented by equation (A.2.3) and it is the same of equation \eqref{41} when $\mathcal{U}_{1}$ is represented by equation (A.3).

\section*{Acknowledgements}
This work has been partially supported by MICINN/FEDER grant
number MTM2011--22587.

\end{document}